# An Asymptotic relation for Hadjicostas Formula

Nikos Bagis
Department of Informatics, Aristotele University
Thessaloniki Greece
bagkis@hotmail.com

**Abstract**
We derive an asymptotic formula which in some cases generalizes Hadjicostas formula.

**Introduction**

Hadjicostas conjecture the following formula [5]

$$\int_0^1\int_0^1 \frac{1-x}{1-xy}(-\log(xy))^s\, dxdy = \Gamma(s+2)\left(\zeta(s+2)-\frac{1}{s+1}\right) \quad :(1)$$

for $\text{Re}(s) > -2$. The formula first proven by Chapman [4].

**The Theorem**

Let $F$ be analytic in $(-a,a)$, $a \geq 1$, converges absolutely in 1 and $F(0) = 0$, with $F(x) = kx^v + rx^{v+1} + mx^{v+2} + \ldots$, $v \in \mathbb{N}$. Then

$$\sum_{k=1}^N F\left(\frac{1}{k}\right) - \int_1^N F\left(\frac{1}{t}\right) dt = \int_0^1\int_0^1 \frac{1-x}{1-xy}\frac{G(-\log(xy))}{\log(xy)^2}dxdy = O\left(\frac{1}{N^v}\right) \quad :(2)$$

where

$$F(s) = \int \frac{1}{s^2}(LG)\left(\frac{1}{s}\right)ds \quad :(3)$$

and $L$ is the Laplace Transform.

We will need from [2] the next

**Lemma.** Let $F$ be a function as in Theorem, then

$$\sum_{k=1}^N F\left(\frac{1}{k}\right) - \int_1^N F\left(\frac{1}{t}\right)dt = c(F) + O\left(\frac{1}{N}\right)$$

where

$$c(F) := F'(0)\gamma + \sum_{s=2}^{\infty}\frac{F^{(s)}(0)}{s!}\left(\zeta(s) - \frac{1}{s-1}\right).$$

**Proof of Theorem.**

We rewrite (1) in the form

$$\int_0^1\int_0^1 \frac{1-x}{1-xy}\frac{(-\log(xy))^{s+1}}{(s+1)!(\log(xy))^2}dxdy = \zeta(s+2)-\frac{1}{s+1}$$

Setting $s \to s-2$, $s$-positive integer, we get after multiplying both sides of (2) with the derivative at zero $G^{(s)}(0)$ of some function $G$ and summing

$$\int_0^1\int_0^1 \frac{1-x}{1-xy}\frac{G(-\log(xy))}{\log(xy)^2}dxdy = G'(0)\gamma + \sum_{s=2}^{\infty} G^{(s)}(0)\left(\zeta(s)-\frac{1}{s-1}\right)$$

Using the Lemma along with the relation $\frac{F^{(m)}(0)}{(m-1)!} = G^{(m)}(0)$, we get the relation between $F$ and $G$ and the proof is complete.

**Examples**

1) Set $G(x) = x$, then

$$\int_0^1\int_0^1 \frac{1-x}{(1-xy)\log(xy)}dxdy = \gamma$$

2) Set $G(x) = x\cos(ax)$, $|a|<1$, then

$$\int_0^1\int_0^1 \frac{(1-x)\cos(a\log(xy))}{(1-xy)\log(xy)}dxdy = \frac{1}{2}\left(\log(1+a^2)-\psi(1-ia)-\psi(1+ia)\right)$$

$\psi$ is the polygamma function (see [1]).

3) Set $G(x) = x^2 J_0(x)$, where $J$ is the Bessel function of the first kind then

$$\int_0^1\int_0^1 \frac{(1-x)J_0(\log(xy))}{1-xy}dxdy = -\frac{1}{\sqrt{2}}+\sum_{k=1}^{\infty}\frac{k}{(k^2+1)\sqrt{k^2+1}}$$

4) Set $G(x) = \frac{x^2}{\cosh(2x)}$, then

$$\int_0^1\int_0^1 \frac{(1-x)\text{sech}(2\log(xy))}{1-xy}dxdy = -\frac{\pi+2\log\left(\frac{\pi}{8}\right)}{2\sqrt{2}}+\frac{1}{32}\sum_{k=1}^{4}\psi'\left(\frac{1}{4}+\frac{k}{8}\right)$$

Another interesting example, involving Catalan's constant is with $G(x) = \frac{x^3}{\cosh(x/2)}$ which gives

$$-\int_0^1\int_0^1 \frac{(1-x)\log(xy)\text{sech}(1/2\log(xy))}{1-xy}dxdy = -8+8C-\frac{1}{4}\psi_3\left(\frac{3}{4}\right)$$

Generalizing the above we have if $G(x) = \dfrac{x^m e^{-cx}}{\cosh(x/2)}$, then

$$-\int_0^1\int_0^1 \frac{(1-x)(xy)^c(-\log(xy))^m \operatorname{sech}(1/2\log(xy))}{(1-xy)\log(xy)^2}\,dxdy =$$

$$= -2^{2-m}\psi_{m-2}\left(\frac{2c+3}{4}\right) + 2^{2-m}\psi_{m-2}\left(\frac{2c+5}{4}\right) - 2^{1-m}\psi_{m-1}\left(\frac{2c+3}{4}\right)$$

and if $G(x) = \dfrac{x^m}{\cosh(cx)}$ then

$$2^{2m-1}c^m \int_0^1\int_0^1 \frac{(1-x)(-\log(xy))^m \operatorname{sech}(c\log(xy))}{(1-xy)\log(xy)^2}\,dxdy =$$

$$= 4c\left(\psi_{m-2}\left(\frac{3c+1}{4c}\right) - \psi_{m-2}\left(\frac{c+1}{4c}\right)\right) - \sum_{k=1}^{2c}\psi_{m-1}\left(\frac{c+k}{4}\right)$$

An example that $F$, $G$ are not analytic in the origin is with $G(x) = x^2 \log(x)$. Then

$$\int_0^1\int_0^1 \frac{(1-x)\log(-\log(xy))}{1-xy}\,dxdy = \gamma + \frac{\pi^2}{6}\left(1 + \log(2) - 12\log(A) + \log(\pi)\right)$$

Where $A$ is Glaisher's constant i.e $\log(A) = \dfrac{1}{12} - \zeta'(-1)$.